\documentclass[sor,jor,floatfix]{revtex4-2}

\usepackage{amsmath,amssymb,bm,color}
\usepackage{graphicx}

\setcitestyle{numbers,square,comma}

\begin{document}


\title{Coupling between long ranged repulsions and short ranged attractions in a colloidal model of zero shear rate viscosity}

\author{Edmund M. Tang}
\thanks{Contributed equally to this work}

\author{Sabitoj Singh Virk }
\thanks{Contributed equally to this work}
\author{Patrick T. Underhill} \altaffiliation{Author to whom correspondence should be addressed. Electronic mail:underp3@rpi.edu}
\affiliation{{Rensselaer Polytechnic Institute, 110 8th St, Troy, New York 12180, USA}}

\date{\today}

\begin{abstract}
In this work, we analyzed an isotropic colloidal model incorporating both short-range sticky attractions and long-range electrostatic repulsions. We computed the zero-shear viscosity and second virial coefficient for a dilute colloidal suspension (i.e., pair interactions only) as a function of the strength of attractions and repulsions. We also developed an analytical approximation that allows us to better understand the coupling of the two types of interactions. The attractions and repulsions contribute to the zero-shear viscosity and second virial coefficient in different ways, leading to cases with the same second virial coefficient but different zero-shear viscosity. The analytical approximation shows that the mechanism of the coupling of interactions is that long-range repulsions can weaken the influence of short-range attractions. This effect alters how repulsions change the zero-shear viscosity. Acting independently, both attractions and repulsions increase the viscosity coefficient of the system. However, when both type of interactions are considered together, repulsions can screen the effect of attractive interactions thereby reducing the viscosity.
\end{abstract}

\maketitle

\section{Introduction}\label{sec_intro}

Models for colloidal dispersions serve as a useful tool to connect microscopic details to overall fluid properties. In spite of simplifications, such models are still able to capture the macroscopic behavior of various materials. Colloidal models often work under the assumption that the dispersed particles are hard spheres to simplify the hydrodynamic calculations. Such models have been applied to protein solutions, despite the ability of proteins to undergo conformational changes when exposed to large pH variations, the addition of denaturants, or strong shear, among other influences. It has been shown that such models can successfully represent the behavior of protein solutions as long as the conditions are maintained to retain the quasispherical shape of the proteins~\citep{connolly_2012, zydney_2015, ross_1977}.

In many studies, the dispersed particles are assumed to possess only a single type of particle-particle interaction. When looking at particle aggregation and self-assembly, the particles are treated as having ‘sticky’ interactions in which the particles reversibly adhere when they come into contact~\citep{baxter_1968, russel_1984, stell_1991, holmes-cerfon_2013, holmes-cerfon_condens}. Such systems are inherently unstable and such studies are often interested in the transition from a collection of singlets to clusters of particles. Even for systems with complex, orientation-dependent attractions, it has been found that the details of the interaction potential are not always important and a simple sticky interaction potential is sufficient to deduce the macroscopic properties of these systems~\citep{holmes-cerfon_2017}. Systems featuring orientation-dependent interactions include protein molecules~\citep{piazza_1998, lenhoff_2017} and patchy microparticles~\citep{hayakawa_2016}.

Stable dispersions are often thought of as an ensemble of particles dominated by repulsive interactions to provide stability. These repulsive interactions may take the form of electrostatic interactions due to the charge on suspended microparticles~\citep{fryling_1963, prieve_1987, prieve_1990, carnero-ruiz_1997}, or steric interactions due to polymeric chains that prevent flocculation~\citep{napper_1977}. Studies involving electrostatic interactions can be modeled using a screened coulomb potential while steric interactions may be modeled using an excluded shell model. However, electrostatic interactions may be simplified to having an equivalent excluded shell distance rather than having a repulsive potential that varies with distance.

Studies involving repulsive interactions often assume that the repulsive effects exclude the possibility of experiencing attractive forces and that these interactions are time invariant.  Such studies are often designed to maintain strongly repulsive interactions that would preclude underlying attractive interactions. For example, studies involving protein solutions are performed far from the isoelectric point, such that strong electrostatic repulsions exist between proteins~\citep{connolly_2012, morbidelli_2015}. However, even stabilized dispersions will eventually aggregate when given the time to degrade~\citep{feke_1984}. London-van der Waals attractions and hydrodynamic interactions become relevant when repulsive forces are weak.

Both attractive and repulsive forces should be considered in any colloidal system wherein the two types of forces are comparable in magnitude. Thus in systems involving reversible aggregation~\citep{rogers_2005}, protein solutions~\citep{pathak_2013, pathak_2015}, or gelation~\citep{fryling_1963, swan_2019}, the interplay between attractive and repulsive forces between colloidal particles or with the substrate may be important. The view that many colloidal systems experience a single interaction is, fundamentally, a simplification. Including both attractions and repulsions will paint a more accurate picture of the rheology of colloidal systems.

In this article, we consider a colloidal model incorporating short-range sticky attractions and long-range electrostatic repulsions. We calculate the low-shear viscosity for a dilute suspension of such sticky electrostatic spheres and demonstrate how the viscosity and second virial coefficient can be varied independently for such systems.

\section{\label{sec_background}Background}

This section provides the details of the system studied. We consider a dilute suspension of rigid, spherical particles and compare the second virial coefficient and zero-shear viscosity. These are calculated as a function of the particle pairwise interaction potential. All energy terms are nondimensionalized using the thermal energy of the system $kT$ as the energetic scale, with $k$ being the Boltzmann constant and $T$ being the absolute temperature of the system. Further, all distances are nondimensionalized using the particle radius $a$ as a length scale.

\subsection{\label{sec_interaction_potential}Interaction Potential}

The macroscopic behavior of colloidal dispersions is determined by the behavior of the underlying microstructure. This in turn is dependent on the interactions between individual colloidal particles. Not only do the interactions between particles yield hydrodynamic stresses, but Brownian motion and the forces between particles act to maintain the equilibrium microstructure, which resists perturbations to the microstructure~\citep{batchelor_1977}. In this section, a pair interaction potential between two rigid hard spheres at dilute concentrations is used that possesses both short-range sticky attractions and long-range electrostatic repulsions. By varying the strength of each, we can understand how the combination determines the overall properties.

To model short-range attraction, the Baxter Sticky Sphere (BSS) model is typically used, wherein an infinitely steep hard-core repulsion is followed by a short-range square-well attraction. The dimensionless pair interaction potential for the BSS model is given by~\citep{baxter_1968, russel_1984}:
\begin{equation} \label{eqn:Vss}
    V_{BSS}\left(\rho\right) =
        \begin{cases}
        \infty & \rho\leq2 \\
        \text{log}\left[6\tau\left(R-2\right)\right] & 2 < \rho < R \\
        0 & R \leq \rho
        \end{cases}
\end{equation}
where $\rho$ is the dimensionless center-center distance between two particles, $\tau$ is the stickiness parameter, and $R$ is the outer bound of the attractive well. Results are taken in the limit of $R \to 2$ and thus quantified by only $\tau$. At dilute concentrations, the equilibrium radial distribution function is given by the Boltzmann distribution of pairs as:
\begin{equation} \label{eqn:gss}
    g_{ss}\left(\rho\right) = \theta\left(\rho-2\right) + \left(\frac{1}{6\tau}\right)\delta\left(\rho-2\right)
\end{equation}
where $\theta$ is the Heaviside function and $\delta$ is the Dirac delta function. The first term is the hard sphere contribution, while the second term is due to the sticky part of the interaction potential. The integral of the second term is $1/(6\tau)$ and can be used to quantify the strength of attraction. The particles are ‘stickier’ with smaller values of $\tau$.

While Baxter's sticky sphere model is commonly used, it can lead to problems in some numerical methods due to the discontinuities in the derivative of the potential. Instead of using the BSS model, we use a Morse potential that is continuous and captures the important features of the BSS model~\citep{holmes-cerfon_2017}. Parameters are chosen to replicate the hard-core repulsion followed by a short-range attractive well similar to that in the BSS model. The Morse (M) potential for $\rho>2$ is described by:
\begin{equation} \label{eqn:morse}
    V_M\left(\rho\right) = \varepsilon_d e^{-\left(\rho-2\right)/\epsilon}\left(e^{-\left(\rho-2\right)/\epsilon}-2\right)
\end{equation}
where $\varepsilon_d$ is the depth of the attractive well, occurring at a distance 2, and $\epsilon$ is the range of the sticky interaction. We calculate an effective stickiness parameter $\tau^*$ for this model using the integral of the equilibrium radial distribution function $g_M$:
\begin{equation} \label{eqn:tauEff}
    \frac{1}{4}\int_{2}^{\infty}\left[g_{M}\left(\rho\right)-1\right]\rho^2d\rho = \frac{1}{6\tau^*}.
\end{equation}

Because the equilibrium radial distribution function is $g_{M}\left(\rho\right)=e^{-V_{M}\left(\rho\right)}$, defining the stickiness using this integral is equivalent to using the second virial coefficient to define the stickiness. Matching second virial coefficients is a common approach for comparing potentials, especially for the Baxter sticky sphere model~\citep{mewis_2011}.

We consider a system where long-range interactions are dominated by isotropic screened electrostatic repulsions. An electrically charged particle in solution develops a diffuse counterion cloud that shields the surface charge. The extent of this electric double layer is characterized by the dimensionless Debye length $1/\kappa$. The charge is quantified by the parameter $\alpha$, which is a dimensionless group comparing the electrostatic force to the Brownian force. Any motion of the bulk fluid will perturb the electric double layer. In response, Brownian and electrostatic forces restore equilibrium by moving ions relative to the fluid. Considering a system with strong electrostatic forces ($\alpha \gg 100$) and electric double layers that are only slightly perturbed from equilibrium, an approximate screened Coulomb (SC) potential is given by~\citep{russel_1976}:
\begin{equation} \label{eqn:Vsc}
    V_{SC}\left(\rho\right)=\alpha e^{-\kappa\rho}/\kappa\rho
\end{equation}
The dimensionless group $\alpha$ is given by:
\begin{equation} \label{eqn:alpha}
    \alpha = \frac{\epsilon_{f}\psi^{2}_{0}a}{kT}\kappa e^{2\kappa}
\end{equation}
where $\epsilon_f$ is the permittivity of the fluid, $\psi_{0}$ is the surface potential and a being the particle radius.

The pair interaction potential for sticky electrostatic spheres (SES) is taken to be the sum of the Morse potential and the screened Coulomb potential (equations $\left(\ref{eqn:morse}\right)$ and $\left(\ref{eqn:Vsc}\right)$):
\begin{equation} \label{eqn:Vses}
    V_{SES}\left(\rho\right) = V_{M}\left(\rho\right) + V_{SC}\left(\rho\right).
\end{equation}

\begin{figure}[t]
    \centering
\includegraphics[scale= 0.9]
{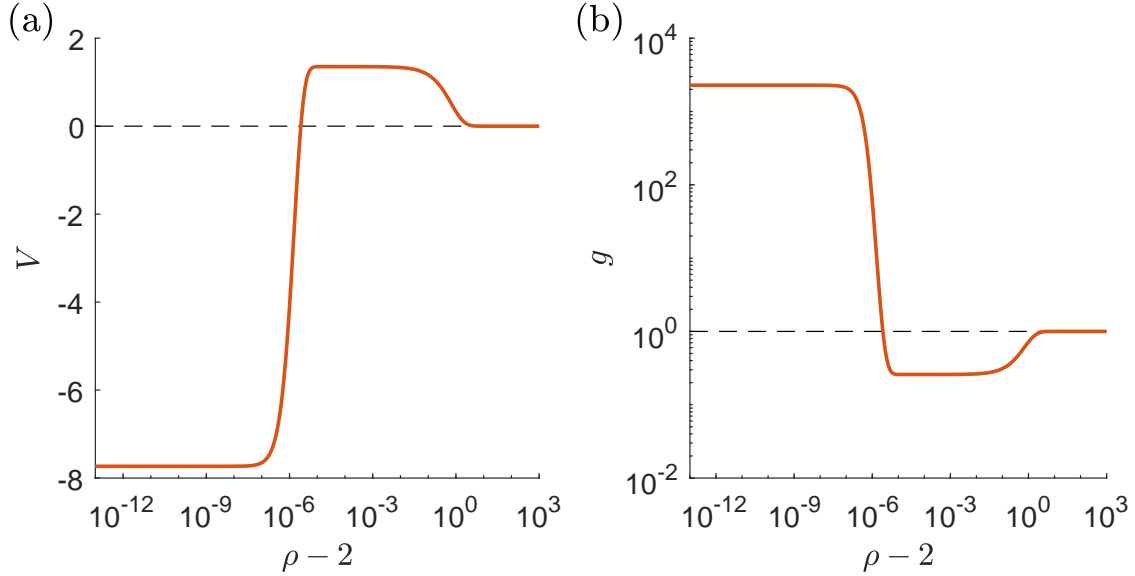}\caption{The interaction potential is plotted against particle separation for a particle with effective stickiness ($1/\tau^* = 0.02$) and electrostatic repulsion ($\alpha = 20$). (b) The equilibrium distribution function is plotted for the same particle. The dashed line represents the hard sphere result.}
    \label{fig:01}
\end{figure}

An example potential and equilibrium distribution $g=e^{-V}$ are shown in figure \ref{fig:01}. In the interaction potential, there is in an attractive well over $2<\rho \leq 2+3\epsilon$ followed by a repulsive peak that gradually decays as $\rho\to\infty$. The corresponding equilibrium distribution is calculated for each interaction potential as the Boltzmann distribution of the interaction potential. The separation of the particle surfaces is plotted on a log scale to show the behavior near contact and far away. It is worth noting that in this figure, the effective stickiness ($1/\tau^* = 0.02$) is relatively small compared to the range of values that were examined ($0\leq 1/\tau^* \leq 20$). Even for small values of stickiness, the depth of the energy well at contact appears large. This manifests itself similarly for the equilibrium distribution, in which $g$ near contact is large. However, the number of particles near contact is related to the integral of $g$, which is not necessarily large even if the $g$ near contact is large. Similarly, significant irreversible clustering of particles or phase separation should occur only when the number of particles near contact is large.

The second virial coefficient can be calculated using the pair interaction potential from:
\begin{equation} \label{eqn:B2}
    B_{2} = -2\pi a^3 \int_{0}^{\infty}\left(e^{-V\left(\rho\right)}-1\right)\rho^2d\rho.
\end{equation}

\begin{equation} \label{eqn:B2*}
    B_2^* =\frac{B_2}{B_{2,HS}} = -\frac{3}{8} \int_{0}^{\infty}\left(e^{-V\left(\rho\right)}-1\right)\rho^2d\rho.
\end{equation}

where, $B_{2,HS}=\frac{16\pi a^3}{3}$ is the hard sphere result and $B_2^*$ is the second virial coefficient normalized by the hard sphere result ($B_2^* = B_2/B_{2,HS}$). The second virial coefficient is often regarded as an indicator of the interactions between particles in colloidal systems. The condition in Fig.~\ref{fig:01} has $B_2^* = 3.75$. The second virial coefficient is effective at capturing the osmotic pressure when examining the combination of different, opposing interactions~\citep{neal_1998}. However, the effects of such interaction on the viscosity of a suspension are not as intuitive or obvious.

\subsection{\label{zero-shear} Zero-Shear Viscosity}

The bulk stress of a suspension of particle arises from three distinct sources~\citep{batchelor_1977, batchelor_1972b, batchelor_1976, russel_1980, russel_book_1987}. The first are hydrodynamic stresses originating from the viscous nature of the fluid. The presence of rigid spheres adds extra stresses to the system, arising from not only the influence of individual particle on the fluid flow, but also from the hydrodynamic interactions between pairs of particles. The hydrodynamic stress is affected by Brownian motion and the interparticle potential through the equilibrium distribution function $g\left(\rho\right)$.
Both the Brownian motion of particles and the interparticle potential contribute to the bulk stress of the system, and comprise the other two sources. When stress is applied to the system, the microstructure is perturbed. These mechanisms resist this perturbation and seek to restore the microstructure. This equilibrium microstructure thus contributes to the stress even at the low-shear limit and causes non-Newtonian behavior at high shear rates.
The microstructure of a dilute suspension of particles can be described using the pair distribution function $P\left(\rho\right)$, which reflects the probability of finding a second particle at distance $\rho$ relative to a test particle. Due to the shear on the system, the pair distribution function will deviate from the equilibrium value of $g\left(\rho\right)$. This deviation is described by the perturbed distribution function $f\left(\rho\right)$. The pair distribution function is related to the equilibrium distribution function and the perturbed distribution function by~\citep{russel_1984}:
\begin{equation} \label{eqn:P}
    P\left(\rho\right) = ng\left(\rho\right)\left[1-\left(3\pi\mu_{0} a^{3}\right)\left(\frac{\bm{\rho}\cdot\bm{E}\cdot\bm{\rho}}{\rho^2}\right)f\left(\rho\right)\right]
\end{equation}

where, $\bm{E}$ is the symmetric rate-of-strain tensor and $\mu_0$ is the viscosity of the Newtonian suspending fluid. In the dilute limit, the pair distribution must satisfy the conservation equation for particles as described by Batchelor. Russel reduced this equation, via a regular perturbation expansion for small shear rate,
to~\citep{russel_1984}:
\begin{equation} \label{eqn:bvpf}
    \frac{d}{d\rho}\rho^{2}G\frac{df}{d\rho}-\rho^{2}G\left(\frac{dV}{d\rho}\right)\frac{df}{d\rho} - 6Hf = -\rho^{2}W + \rho^{3}\left(1-A\right)\frac{dV}{d\rho}
\end{equation}

with boundary conditions
\begin{equation} \label{eqn:bvpf_bc}
    \begin{alignedat}{3}
        \rho\left(1-A\right) + G\frac{df}{d\rho} &= 0\quad          &\text{at}\quad \rho &= 2 \\
        f    &\to 0\quad   &\text{as}\quad \rho &\to \infty
    \end{alignedat}
\end{equation}
This equation is valid for an arbitrary isotropic interaction potential $V$. Solving this boundary value problem involves a set of known hydrodynamic functions ($G$, $H$, $A$, $B$, $W$) for a pair of equal rigid spheres. These functions depend only on $\rho$, and have defined asymptotic forms in the limits of wide separation ($\rho\to\infty$) and near contact ($\rho\to 2$);
\begin{equation} \label{eqn:hydro}
        \begin{alignedat}{1}
        G\left(\rho\right) &=
            \begin{cases}
            1 - \frac{3}{2}\rho^{-1} + \rho^{-3} + \frac{15}{4}\rho^{-4} + \ldots &\rho\to\infty \\
            2\left(\rho-2\right) - \frac{9}{5}\left(\rho-2\right)^{2}\text{log}\left(\rho-2\right)^{-1} - 4\left(\rho-2\right)^{2} + \ldots &\rho\to2
            \end{cases}
        \\
        H\left(\rho\right) &=
            \begin{cases}
            1 - \frac{3}{4}\rho^{-1} - \frac{1}{2}\rho^{-3}+\ldots
           &\rho\to\infty\\
            0.402 + \frac{0.547}{1.349 + \text{log}\left(\rho-2\right)^{-1}} - \frac{0.151}{4.69 + \text{log}\left(\rho-2\right)^{-1}} + \ldots &\rho\to2
            \end{cases}
        \\
        A\left(\rho\right) &=
            \begin{cases}
            5\rho^{-3} - 8\rho^{-5} + \ldots &\rho\to\infty \\
            1 - 4.077\left(\rho-2\right) + \ldots &\rho\to2
            \end{cases}
        \\
        W\left(\rho\right) &=
            \begin{cases}
            75\rho^{-6} + \ldots &\rho\to\infty\\
            6.372 + \ldots &\rho\to2
            \end{cases}
        \end{alignedat}
\end{equation}
In a later section we describe the numerical methods used to solve this boundary value problem and use the solution to compute the viscosity.

After solving for the microstructure of the system, the zero-shear viscosity of the dilute suspension of particles is computed. Batchelor had found the zero-shear viscosity of a dilute suspension of hard spheres, up to order $\phi^2$, to be:
\begin{equation} \label{eqn:mu}
    \mu = \mu_{0}\left(1 + 2.5\phi + c_2\phi^2 + \ldots\right)
\end{equation}
where $\phi$ is the volume fraction of the colloidal particles. The first two terms represent the fluid viscosity and the contribution of isolated particles to the viscosity. The third term $c_2$ or the quadratic coefficient of viscosity, reflects the contribution from the interaction between pairs of particles. For a suspension of hard spheres, Batchelor found that $c_{2,HS} = 6.2$, but later work refining the calculation found that $c_{2,HS} = 5.9$~\citep{mewis_2011, cichocki_1993, bergenholtz_2002}. In other works where the relative viscosity is expressed as a function of the weight concentration of colloidal particles and intrinsic viscosity, the coefficient to the second order term is the Huggins coefficient $k_H$, where $k_H=c_{2}/2.5^2$. We are interested in how the sticky electrostatic interactions manifest in the quadratic coefficient.

Previous work has found that the quadratic coefficient is the sum of a hydrodynamic contribution $c_2^H$, a Brownian contribution $c_2^B$ and an interaction contribution $c_2^I$ and can be computed via the following integrals~\citep{batchelor_1977,russel_book_1987}:
\begin{equation}
\label{eqn:c2}
    \begin{alignedat}{1}
        c_{2} &= c_{2}^{H} + c_{2}^{B} + c_{2}^{I} \\
        c_{2}^{H} &= \frac{5}{2}\left[1+3\int_{2}^{\infty}J\left(\rho\right)g\left(\rho\right)\rho^{2}d\rho\right] \\
        c_{2}^{B} &= \frac{9}{40}\left[\int_{2}^{\infty}\rho^{2}W\left(\rho\right)g\left(\rho\right)f\left(\rho\right)d\rho\right] \\
        c_{2}^{I} &= \frac{9}{40}\left[\int_{2}^{\infty}\rho^3\left(1-A\left(\rho\right)\right)\frac{dg}{d\rho}f\left(\rho\right)d\rho\right]
    \end{alignedat}
\end{equation}
The hydrodynamic contribution encompasses stresses that arise due to the flow of fluid around each particle. Additional stresses arise from a particle being entrained in the flow caused by another particle and dependent on the distance between particles. Near-field hydrodynamic interactions are accounted for by $J\left(\rho\right)$, which decays as $J\sim\frac{15}{2}\rho^{-6}$ as $\rho\to\infty$. The Brownian contribution arises from thermodynamic forces restoring the microstructure to its equilibrium after a perturbation. The interaction contribution covers the resistance to perturbation that arises from the force directly exerted on a particle due to another particle.

\section{\label{sec_results_discussion} Results and Discussion}

\subsection{\label{sec_numerical solution}Numerical Solution}
Numerical calculations were performed in MATLAB. The interaction potential is specified through four parameters. For the sticky part of the interaction potential (equation $\left(\ref{eqn:morse}\right)$), the depth of the attractive well was chosen within $0\leq\varepsilon_d\leq 63$, while the range of sticky interaction was chosen to be small to simulate the sticky limit~\citep{holmes-cerfon_2017}. A value of $\epsilon=10^{-6}$ was chosen as it was the smallest value before encountering machine precision issues. For the electrostatic part (equation $\left(\ref{eqn:Vsc}\right)$), the strength of electrostatic repulsion was chosen within $0\leq\alpha\leq30$, while the Debye length was fixed $\kappa^{-1} = 1$.

The hydrodynamic functions ($A$, $G$, $H$, $J$, $W$) utilize the known behavior near-contact and far-from-contact wherever possible. For simplicity, the intermediate values of the functions were approximated by interpolating Batchelor’s tabulated values~\citep{batchelor_1972, batchelor_1972b}. This results in a $c_{2,HS}$ that most closely aligns with Batchelor’s original result. Although Batchelor's result was refined in later works using more complex numerical approaches, they do not affect the conclusions of this work.

The boundary value problem described in equation $\left(\ref{eqn:bvpf}\right)$ was solved using the MATLAB function $bvp4c()$ for different interaction potentials. This boundary value problem is divergent at contact ($\rho=2$), so a Neumann boundary condition was chosen at $\rho = 2+10^{-13}$. Noting that the machine precision of MATLAB is $2.2204\times10^{-16}$, this is the closest to contact that the boundary value problem could be evaluated before breaking down. The far boundary condition was a Dirichlet boundary condition evaluated at $\rho=1002$, a point far enough away to be indistinguishable from $\rho\to\infty$. The solution of the boundary value problem for $f\left(\rho\right)$ was then used to compute the integrals in equation $\left(\ref{eqn:c2}\right)$ to obtain $c_2^H$, $c_2^B$, and $c_2^I$. These integrals were computed using MATLAB’s $integral()$ function.

\begin{figure}[b]
    \centering
\includegraphics[scale=0.9]{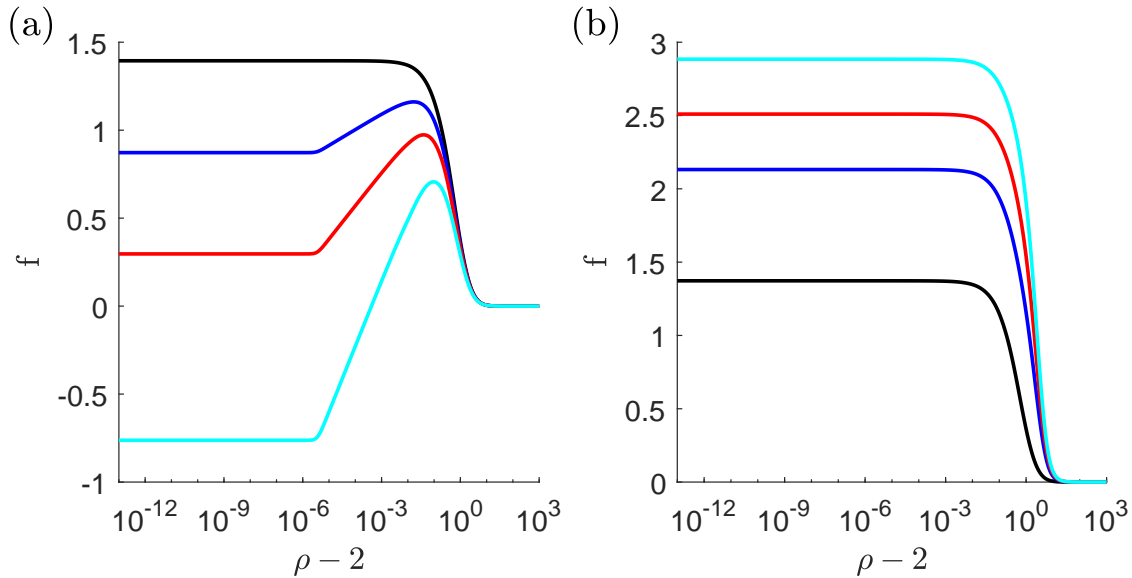}

    \caption{(a) The perturbed distribution function is plotted against particle separation for particles with varying stickiness and no electrostatic repulsion. The black is for a hard sphere ($1/\tau^* = 0$), the blue is $1/\tau^* = 0.2$, the red is $1/\tau^* = 0.5$, and the cyan is $1/\tau^*=1.5$. For all curves, $\epsilon = 10^{-6}$. (b) The perturbed distribution function plotted against particle separation for particles with no stickiness and varying strength of electrostatic repulsion. The black is for a hard sphere ($\alpha = 0$), the blue is $\alpha = 12$, the red is $\alpha = 18$, and the cyan is $\alpha = 24$. For all curves, $\kappa^{-1} = 1$.}
    \label{fig:02}
\end{figure}

Figure \ref{fig:02} (a) shows the perturbed distribution function for a purely sticky particle for varying stickiness without electrostatic repulsion. The distribution function is reduced relative to the hard-sphere case, with stronger attractions leading to more reduction. Recall from equation $\left(\ref{eqn:bvpf}\right)$  that $f$ represents how the distribution is changed due to weak flows. Negative values of $f$ mean that there are more particles than equilibrium at certain angles relative to the flow. For $\rho-2 < \epsilon = 10^{-6}$, $f$ is nearly constant. The deviation of $f$ from the hard-sphere result extends to large enough distances where the Morse potential has decayed to zero. In this region, the reduction relative to the hard-sphere result behaves nearly logarithmically and vanishes near $\rho-2\approx 1$. The work of Cichocki \& Felderhof \cite{cichocki_1990} also showed a logarithmic behavior of $f$ for the Baxter model.

 Based on the numerical results in Figure 2 (a), 2(b) and additional results with larger $\epsilon$ (data not shown) the contact value, $f_C$, given by equation $\left(\ref{eqn:f_C}\right)$ in appendix has been approximated. To further analyse $f_C$ with attractions-only contribution (sticky spheres), equation $\left(\ref{eqn:f_C}\right)$ can be substituted with $\alpha$=0 and approximated as $f_{C,SS}$ ,where
\begin{equation} \label{eqn:finsidewell}
    f_{C,SS} \approx 1.37 + \frac{0.21\left(\frac{1}{\tau^*}\right) \ln \epsilon }{-0.05\left(\frac{1}{\tau^*}\right)\ln \epsilon +1}
\end{equation}
In the limit of zero sticky interactions, $\frac{1}{\tau^*} \to 0$, the value of $f_{C,SS}$ corresponds to the hard-sphere case. In the limit of strong attractions, $f_{C,SS}$ approaches a limiting value of $-2.83$, a similar behavior observed by Cichocki \& Felderhof~\citep{cichocki_1990} for highly sticky particles (in their work denoted as $f_w$ = -2.963).

For sticky attractions, it is common to examine the limit where the range of interaction approaches zero, which here corresponds to $\epsilon \to 0$. Because of the logarithm, this limit is slow to converge. Even for a small value of $\epsilon=10^{-6}$ used here, $f_{C,SS}$ only approaches the limiting value for $\frac{1}{\tau^*}$ much greater than $3$. In practice, the sticky attraction model is often used to represent phenomena such as hydrophobic attractions within water or depletion attractions. Although the range of these interactions is often less than the size of the objects/spheres, using the limiting value of $f_{C,SS}$ could lead to significant errors.

Figure \ref{fig:02} (b) shows the perturbed distribution function for varying strength of electrostatic repulsion, with no attractions. Electrostatic repulsion increases the perturbed distribution function with increasing electrostatic repulsion. For a Debye length of $\kappa^{-1} = 1$, the effect of electrostatic repulsion extends somewhat beyond the Debye length, going as far as $\rho-2\approx 10$. $f_C$ with electrostatic repulsions-only contribution (electrostatic spheres) can be approximated from  equation $\left(\ref{eqn:f_C}\right)$ by  \begin{equation} \label{eqn:felectrostatics at contact}
    f_{C,ES}=f_C\left(\frac{1}{\tau^*}=0\right) \approx\frac{2}{3}\alpha e^{-2\kappa}+ 1.37
\end{equation}

Using the equilibrium distribution function and the perturbed distribution function for a particular set of parameters, we are able to compute the quadratic coefficient of the viscosity, $c_{2}$. In order to compare with the results for the combination of interactions, we first show the results for the individual interactions and how the numerical results compare with previous analysis in the literature.

For Baxter sticky hard spheres, Russel originally calculated the viscosity under the assumption that the perturbed distribution function does not vary with varying stickiness~\citep{russel_1984}. Cichoki \& Felderhof later found an error in Russel’s calculations. The corrected quadratic coefficient is~\citep{cichocki_1990}:
\begin{equation} \label{eqn:c2Russel}
    c_{2,SS}=5.931+0.76/\tau
\end{equation}

\begin{figure}[b]
    \centering
    \includegraphics[scale = 0.9]{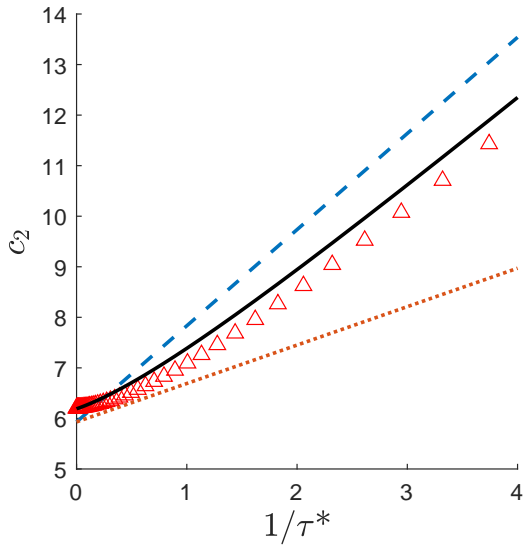}
    \caption{The quadratic coefficient of viscosity is plotted against the strength of stickiness ($1/\tau^*$) for a purely sticky particle (triangles). The dashed line is Cichocki’s theory (equation $\left(\ref{eqn:c2Cichocki}\right)$) while the dotted line is Cichocki’s corrected version of Russel’s theory (equation $\left(\ref{eqn:c2Russel}\right)$). The solid line represents the analytical solution (equation $\left(\ref{eqn:c2SES_SSAnalytical}\right)$).}
    \label{fig:03}
\end{figure}

They further point out that Russel’s original assumption of an unchanging perturbed distribution function to be incorrect. When accounting for a changing $f$, Cichocki \& Felderhof found the quadratic coefficient for the Baxter sticky sphere model to be
\begin{equation} \label{eqn:c2Cichocki}
    c_{2,SS} = 5.931 + 1.899/\tau
\end{equation}

In figure \ref{fig:03}, our numerical results using the Morse potential are compared against the theories of the Baxter model of Russel~\citep{russel_1980, cichocki_1990} and Cichocki \& Felderhof~\citep{cichocki_1990}. We find that our numerical solution follows more closely to the corrected Russel curve for stickiness less than one ($1/\tau^* < 1$), while it follows more closely the theory of Cichocki for high value of stickiness ($1/\tau^* > 1$). We can better understand this using an analytical approximation of our numerical solution. The general formula is given in the Appendix. For the case of no electrostatic repulsions, it becomes

\begin{equation} \label{eqn:c2SES_SSAnalytical}
    c_{2,SS} = 6.19 + \frac{1}{\tau^*}\left(1.11 - 0.26f_{C,SS}\right) = 6.19 + \frac{1}{\tau^*}\left(0.75+\frac{-0.055\left(\frac{1}{\tau^*}\right) \mathtt{\ln{\epsilon }}}{-0.05\left(\frac{1}{\tau^*}\right)\mathtt{\ln{\epsilon}}+1}\right)
\end{equation}

Note that we use approximate values to the hydrodynamic functions from Batchelor, resulting in a constant term in $c_2$ that is more similar to earlier calculations of $c_{2,HS}$ rather than following Cichocki's~\citep{cichocki_1988_diff, cichocki_1988_susp} or other later results~\citep{mewis_2011}. Because we use the analytical approximation of the hydrodynamic functions at small and large separations, the small difference in $c_2$ is likely from interpolation at intermediate distances.

The values of $c_2$ in Fig.~\ref{fig:03} are monotonically increasing with increasing stickiness, similar to theories of the Baxter model. Bergenholtz \& Wagner~\citep{bergenholtz_1994} examined the square-well potential and showed that $c_2$ could decrease slightly before increasing. This effect was shown to be due a non-monotonic perturbed distribution function $f\left(\rho\right)$. Our analytical approximation for the perturbed distribution function $f_{C,SS}$ (equation $\left(\ref{eqn:finsidewell}
\right)$) is monotonically increasing with increasing attraction strength. The difference could be due to the much smaller range of interaction used in our work and the difference between the Morse potential and the square well potential.

\begin{figure}[b]
    \centering
    \includegraphics[scale = 0.9]{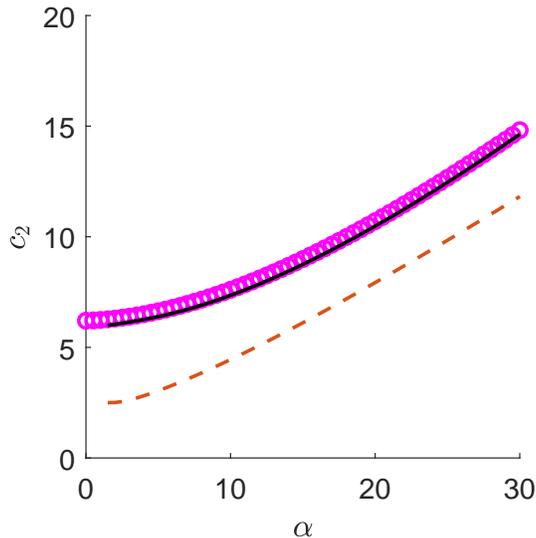}
    \caption{The quadratic coefficient of viscosity plotted against the strength of electrostatic repulsion for a purely repulsive particle (circles). The dashed line is a theoretical curve for highly repulsive particles ($\alpha\gg1$) from equation $\left(\ref{eqn:c2ES}\right)$. The solid line represents an analytical solution for sticky electrostatic spheres from equation $\left(\ref{eqn:c2SES_SC}\right)$ with a $A_{1}$ of 2.35 from equation $\left(\ref{eqn:c2ISES_SC}\right)$ .}
    \label{fig:04}
\end{figure}

We can also look at the case with only electrostatic repulsions and zero sticky attractions, and compare with previous results by Russel~\citep{mewis_2011,russel_1976}. They found the quadratic coefficient to be
\begin{equation} \label{eqn:c2ES}
    c_{2,SC}=\frac{5}{2}+\left(\frac{3}{40}\right)\left(\frac{1}{\kappa}\ln\frac{\alpha}{\ln\left(\alpha/\ln\alpha\right)}\right)^5
\end{equation}

This theory was developed under the assumption that the particles were strongly repulsive ($\alpha\gg 1$) and has been shown to successfully represent the results of experiments for strong repulsions.

Figure \ref{fig:04} compares our numerical results with this analytical approximation. Note that Russel's theory does not approach the hard sphere result as electrostatic repulsion decreases ($\alpha\to 0$). Our analytical approximation from the Appendix can be applied to this case without sticky attractions and obtain
\begin{equation} \label{eqn:c2SES_SC}
\begin{split}
    & c_{2,SC}=
    \frac{5}{2}+ 2.03 e^{-V_C}+1.66+ \\ & A_{1}\left(\frac{3}{40}\right)\left(\frac{1}{\kappa}\ln\frac{\alpha}{\ln\left(\alpha/\ln\alpha\right)}\right)^2{\ln\left(\frac{\alpha}{\ln\alpha}\right)}\left(\frac{1}{6}\left(\frac{1}{\kappa}\ln\frac{\alpha}{\ln\left(\alpha/\ln\alpha\right)}\right)^2{\ln\left(\frac{\alpha}{\ln\alpha}\right)}+1.37\right)
    \end{split}
\end{equation}

It contains the same physical features as Russel's result and same qualitative behavior. However, because it matches the numerical results better in this range of $\alpha$, it is useful to use when examining the case with both electrostatic repulsions and sticky attractions.

Having shown the behavior for exclusively sticky particles and for exclusively electrostatically repelling particles, we next describe the results for the viscosity and second virial coefficient for particles that are simultaneously sticky and repulsive. The stickiness is shown in the range of $0\leq1/\tau^*\leq 20$ and strength of electrostatic repulsion is shown in the range of $0\leq\alpha\leq 30$.

\begin{figure}[h]
    \centering
\includegraphics[scale = 0.9]{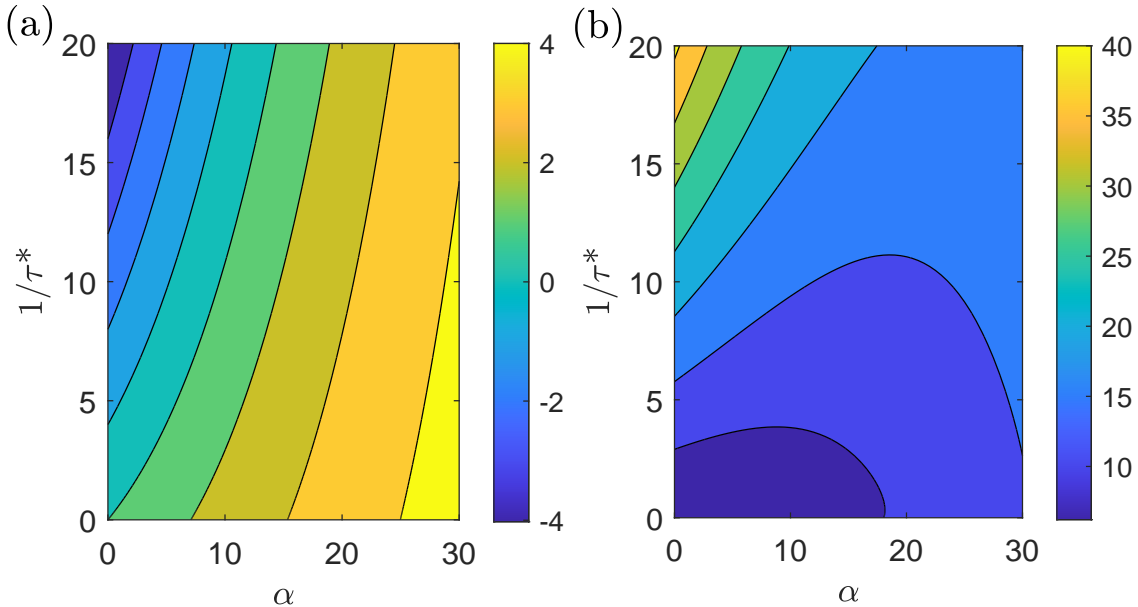}
    \caption{A contour plot of $B_2^*$ (a) and $c_2$ (b) plotted against the stickiness ($1/\tau^*$) and the strength of electrostatic repulsion ($\alpha$).}
    \label{fig:05}
\end{figure}

In figure \ref{fig:05} (a), we plot the contours of the second virial coefficient normalized by the hard sphere result ($B_2^* = B_2/B_{2,HS}$). In figure \ref{fig:05} (b), we plot the contours of the quadratic coefficient ($c_2$). Following either contour plot along y-axis yields the exclusively sticky cases while following along the x-axis yields the exclusively repulsive cases.

In figure \ref{fig:05} (a), we find that increasing stickiness always acts to decrease $B_{2}$, while increasing the electrostatic repulsion always acts to increase $B_{2}$. This trend is independent of the strength of the other interaction, and results in contour curves that are monotonic.

In figure \ref{fig:05} (b), we find that both stickiness and electrostatic repulsion overall act to increase $c_2$ when the other interaction is weak. However, they can also interfere with each other, leading to a weakened increase in viscosity for some combinations of parameters. This results in contour curves that are not monotonic. This is most evident in the middle of the contour plot (for $1/\tau^* = 10$) where the viscosity is lowest for intermediate values of $\alpha$ ($15 < \alpha < 25$).

\begin{figure}
    \centering
    \includegraphics[scale = 0.9]{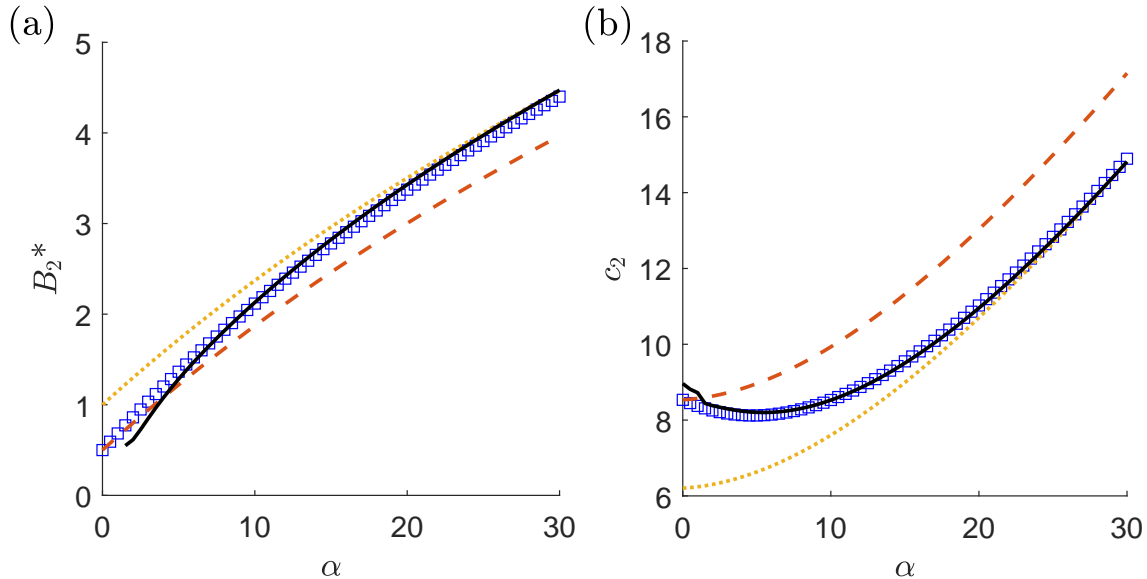}
    \caption{The second virial coefficient (a) and the quadratic coefficient of viscosity (b) plotted against the strength of electrostatic repulsion. The blue squares represent numerical $B_2^*$ and $c_2$ for particles with stickiness $1/\tau^* = 2$. The solid lines represent the analytical approximation from the Appendix. The dashed lines represent the result from additive contributions $\hat{B}_2$ and $\hat{c}_2$ (see equation \ref{eqn:indep}) for the same stickiness. The dotted lines represent $B^*_2$ and $c_2$ for a particle with no stickiness ($1/\tau^* = 0$).}
    \label{fig:06}
\end{figure}

We can quantify the coupling between the stickiness and repulsion by looking at what the behavior would be if the two types of interactions contributed independently to $B_2$ or $c_2$. We can define the two functions
\begin{equation} \label{eqn:indep}
    \begin{alignedat}{1}
        \hat{B}_2\left(\alpha,1/\tau^*\right) &= B^*_2\left(0,1/\tau^*\right) + B^*_2\left(\alpha,0\right) - B^*_2\left(0,0\right) \\
        \hat{c}_2\left(\alpha,1/\tau^*\right) &= c_2\left(0,1/\tau^*\right) + c_2\left(\alpha,0\right) - c_2\left(0,0\right).
    \end{alignedat}
\end{equation}
The results from attractions only and repulsions only are added together. The hard-sphere case must be subtracted so that the overall behavior produces the correct result when only one type of interaction is present. In figure \ref{fig:06}, these sums are compared against the numerical results for the same stickiness $\left(1/\tau^* = 2\right)$ and a range of repulsive strength $\left(0\leq\alpha\leq30\right)$. Curves representing the repulsion-only case (no attractions) are also included, and are a constant shift compared to equation~$\left(\ref{eqn:indep}\right)$. Finally, we compare the numerical results with the approximation developed in the Appendix.

The behavior of the second virial coefficient and viscosity coefficient have a number of many similarities. At $\alpha=0$, they match with equation~$\left(\ref{eqn:indep}\right)$ by construction since only attractions are present. When repulsive interactions are introduced to the system ($\alpha>0$), the numerical results deviate from equation~$\left(\ref{eqn:indep}\right)$ and approach the result including only repulsions. That is, strong enough electrostatic repulsion reduces the impact that stickiness has on both the second virial coefficient or the viscosity coefficient. However, this feature leads to different overall behavior.

For the second virial coefficient, attractions alone lead to a decrease while repulsions alone lead to an increase. Therefore, when repulsions reduce the impact of attractions, it also leads to a increase in the second virial coefficient. This is observed in Figure \ref{fig:06}(a) by the numerical results increasing faster than equation~$\left(\ref{eqn:indep}\right)$.

For the viscosity coefficient, attractions and repulsions alone lead to an increase. However, when repulsions reduce the impact of attractions, it actually leads to a decrease in $c_2$. This occurs in Figure \ref{fig:06}(b) for $0<\alpha<5$. For larger $\alpha$, the repulsions suppress nearly all of the contribution from attractions. This leads to a increase in $c_2$ with increasing $\alpha$ as if there were no attractions. This comparison and analysis is shown explicitly at a relatively small value of the stickiness. This is done because it highlight that the nonmonotonic behavior for the viscosity coefficient occurs even for small attractions for which the second virial coefficient is positive.

Because the analytical approximation developed in the Appendix matches well with the numerical results, the analytical approximation can be used to understand the mechanism of the coupling between the attractions and repulsions. The integral formulas for $B_2$ and $c_2$ add the contributions from pairs of particles including the frequency of finding pairs with a particular $\rho$ and the contribution from pairs at that separation.

The interaction potential between pairs of particles determines the frequency that particles are able to approach certain distances relative to other particles. This is quantified at equilibrium using $g$ and the change in flow using $f$. Attractive interactions increase the frequency with which particles approach each other, while repulsions decrease that frequency. As shown in the Appendix, the equilibrium Boltzmann factor near contact is affected by the repulsive potential near contact. Since the impact of the repulsions depends on the frequency of pairs coming close together, the repulsions can reduce the contribution from the attractions.

At large enough $\alpha$, the repulsions suppress not just the near contact occurrences, but also exclude larger separations. In these conditions, the excluded shell model becomes more accurate. This leads to an increase in $B_2$ and $c_2$ at larger $\alpha$. The decay of the repulsions with increasing $\rho$ means that the repulsions can have an impact near contact for smaller $\alpha$ than is necessary to exclude particles to larger distances. The decrease for $\alpha<5$ in Figure \ref{fig:06}(b) occurs because the repulsions are able to suppress the contribution from the attractions without significantly increasing the $c_2$ directly.

\begin{figure}[h]
   \centering
    \includegraphics[scale=0.9]{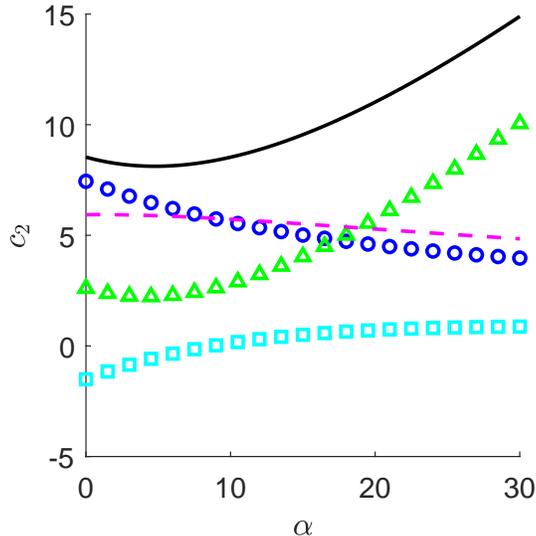}
    \caption{The quadratic coefficient of viscosity contributions for a particle with stickiness ($1/\tau^* = 2$) plotted against the strength of electrostatic repulsion: solid line, total $\left(c_{2}\right)$; $\bigcirc$, hydrodynamic   $\left(c_{2}^{H}\right)$; $\square$,    Brownian $\left(c_{2}^{B}\right)$; $\triangle$, inter-particle interaction $\left(c_{2}^{I}\right)$. For reference is shown in dashed line, $\left(c_{2}^{H}+c_{2}^{B} \right)$.}
    \label{fig:07}
\end{figure}

We can better understand the mechanism by examining the three contributions to $c_2$ in Fig.~\ref{fig:07} for the same conditions as Fig.~\ref{fig:06}. The hydrodynamic contribution monotonically decreases while the Brownian contribution monotonically increases. The sum of these two contributions is shown as a dashed curve, and is nearly constant. The interaction contribution shows a nonmonotonic behavior that nearly parallels the overall sum; the qualitative behavior comes from the inter-particle interaction part alone.

This mechanism appears to be separate from the previous work by Khair \& Brady~\citep{khair_2006}. They examined the microviscosity measured in a system with excluded-shell repulsions. A minimum in microviscosity was observed due to a decrease in hydrodynamic part and increase in interaction part. The minimum occurred where the two contributions are equal. In the system examined here, the location of the minimum in the overall response corresponds more closely to the minimum in the interaction part alone.

\begin{figure}
    \centering
    \includegraphics[scale = 0.9]{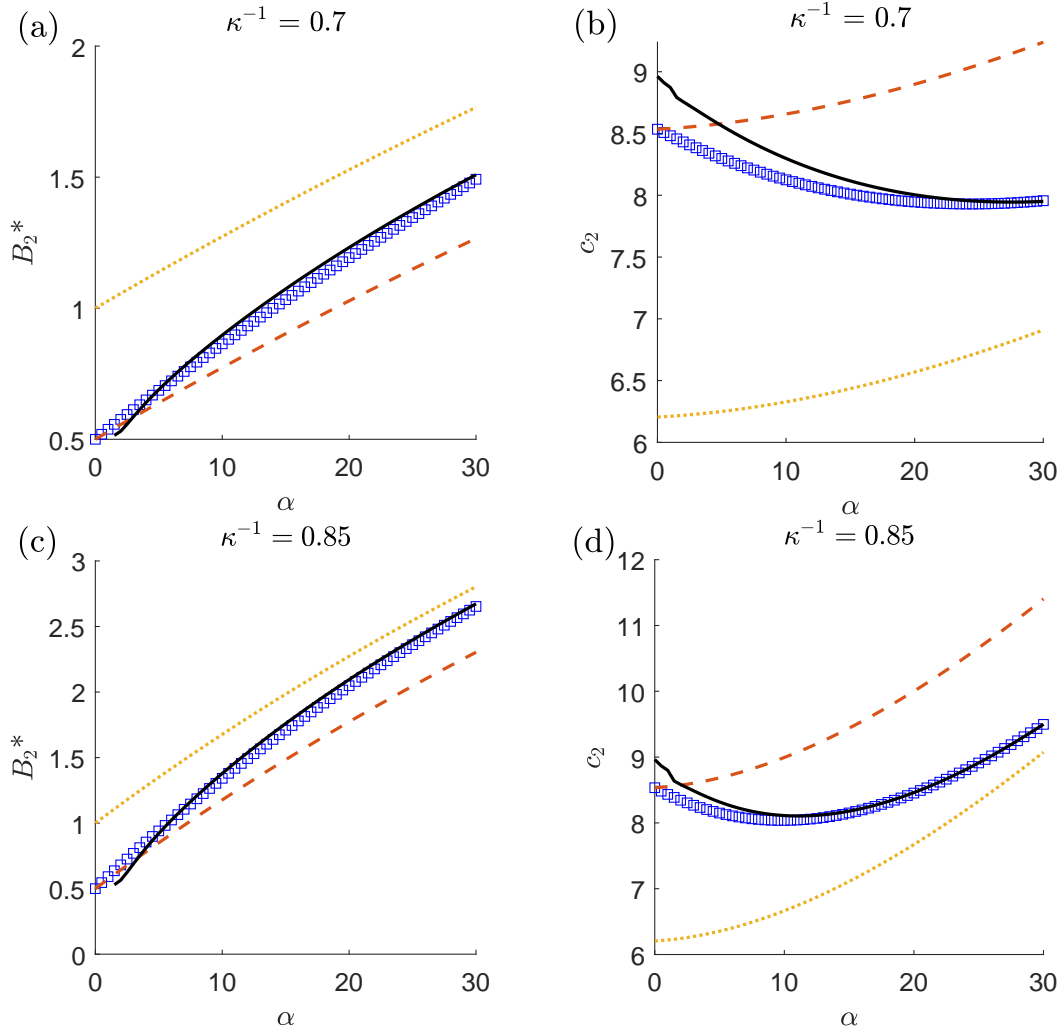}
    \caption{The second virial coefficient (a)(c) and the quadratic coefficient of viscosity (b)(d) plotted against the strength of electrostatic repulsion for Debye lengths \textless 1. The blue squares represent numerical $B_2^*$ and $c_2$ for particles with stickiness $1/\tau^* = 2$. The solid lines represent the analytical approximation from the Appendix. The dashed lines represent the result from additive contributions $\hat{B}_2$ and $\hat{c}_2$ (see equation \ref{eqn:indep}) for the same stickiness. The dotted lines represent $B^*_2$ and $c_2$ for a particle with no stickiness ($1/\tau^* = 0$).}
    \label{fig:08}
\end{figure}

The overall effect of the screened electrostatic repulsions depends on both $\alpha$ and the Debye length $\kappa^{-1}$. All numerical results have been shown for $\kappa^{-1}=1$, while the analytical approximation was written for any $\kappa^{-1}$. Figure \ref{fig:08} shows the numerical results for two values of $\kappa^{-1}<1$. This corresponds to more screening of the repulsions.

The behavior is similar to that seen for $\kappa^{-1}=1$. The weakening of the repulsions from the screening is such that the range $0<\alpha<30$ shown corresponds to relatively small $\alpha$. In this range, repulsions primarily act to indirectly reduce the contribution of attractions. Another important feature visible in these results is that repulsions can act to reduce $c_2$ even when $B_2^{*}$ is greater than $1$. This highlights that the way attractions and repulsions combine together is different for $c_2$ and $B_2^{*}$.

\begin{figure}[h]
   \centering
    \includegraphics[scale=0.9]{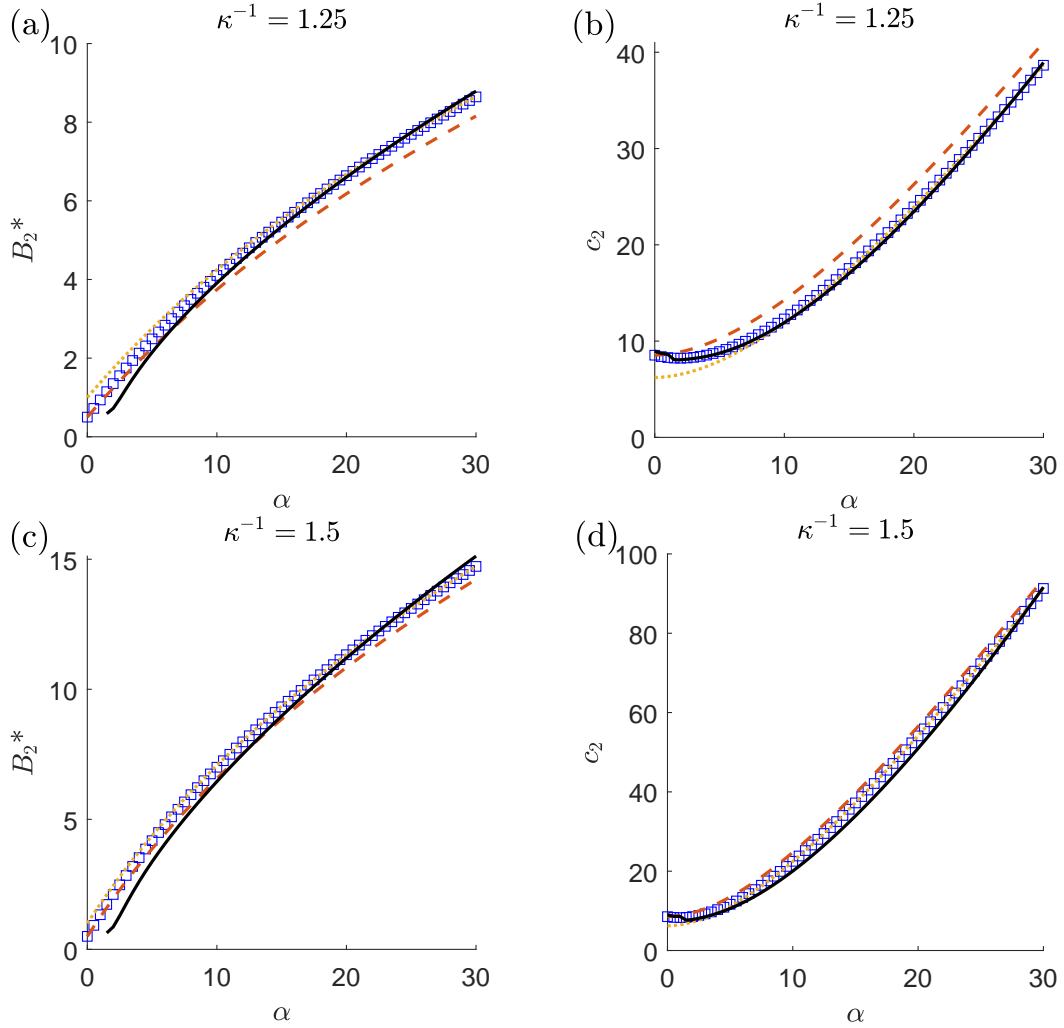}
    \caption{The second virial coefficient (a)(c) and the quadratic coefficient of viscosity (b)(d) plotted against the strength of electrostatic repulsion for Debye lengths \textgreater 1. The blue squares represent numerical $B_2^*$ and $c_2$ for particles with stickiness $1/\tau^* = 2$. The solid lines represent the analytical approximation from the Appendix. The dashed lines represent the result from additive contributions $\hat{B}_2$ and $\hat{c}_2$ (see equation \ref{eqn:indep}) for the same stickiness. The dotted lines represent $B^*_2$ and $c_2$ for a particle with no stickiness ($1/\tau^* = 0$).}
    \label{fig:09}
\end{figure}

Figure \ref{fig:09} shows the numerical results for two values of $\kappa^{-1}>1$. This corresponds to less screening of the repulsions. Because there is less screening, the influence of the repulsions becomes significant even at smaller $\alpha$. Although the repulsions still have both a direct effect and indirect effect on the attractions, the overall strength of repulsions makes the role of attractions less visible.

\section{\label{conclusions} Conclusions}

The values of $B_2$ and $c_2$ can be used to understand how the osmotic pressure and viscosity of dilute suspensions are dependent on the interactions between particles. In order to understand the coupling between attractions and repulsions, we computed the response to two prototypical interactions: sticky attractions using the Morse potential and screened Coulomb repulsions. By comparing the numerical results with an analytical approximation, we were able to understand the mechanism by which the interactions couple: the overall behavior is not the sum of the contributions with attractions and repulsions alone.

When acting alone, sticky attractions act to decrease the second virial coefficient while increasing the viscosity coefficient. When acting alone, screened electrostatic repulsions act to increase both. However, when occuring simultaneously, repulsions can also lead to a decrease in the viscosity coefficient. The physical mechanism is that the repulsions can reduce the likelihood of particles coming in close contact, which reduces the contribution from the sticky attractions. This occurs even when the $B_2$ is positive, such that repulsions are stronger than attractions in their impact on the osmotic pressure.

Because the physical picture underlying this coupling is not specific to the details of the interactions, we expect it to also occur in other situations. For example, many real systems will not have perfectly isotropic interactions or will have interactions of a different functional form. Proteins solutions, for instance, have been shown to experience SALR (short range attraction plus long range repulsion) interactions at low to moderate ionic strengths. At such ionic strengths, the attractions and repulsions have comparable strengths. Tuning the attractions or repulsions with solution conditions or protein mutations may help to reduce the protein solution viscosity. This is an active area of research towards controlling the viscosity of protein formulations and of significant importance to the bio-pharmaceutical and patient healthcare community.

\section*{Acknowledgements}
We gratefully acknowledge support from NSF Grant No. CBET 1803497.

\appendix

\section{\label{sec_analytical solution}Analytical Approximation}
An analytical approximation will be used to better understand the numerical results. For the analytical solution, two features of the Screened-Coulomb potential $V_{SC}$ are particularly important: the potential at/near contact $V_C = V_{SC}|_{\rho=2}$ and the separation distance $L_0$ at which the electrostatic potential balances out with thermal energy~\citep{russel_1976}. This distance is defined as
\begin{equation} \label{eqn:Vsc|L_0}
    \alpha\frac{e^{-\kappa L_{0}}}{\kappa L_{0}} \approx 1
\end{equation}
For $\alpha\gg1$, this can be approximated using the Lambert function to be~\citep{russel_1976}
\begin{equation} \label{eqn:L_0}
    L_0 \approx \frac{1}{\kappa}\ln\frac{\alpha}{\ln\left(\alpha/\ln\alpha\right)}
\end{equation}
This is not a good approximation for smaller $\alpha$.
For $0<\alpha<1.5$,
\begin{equation} \label{eqn:L_0verysmallalpha}
    L_0 \approx \frac{1}{\kappa}\frac{\alpha}{\exp\left(\alpha/\exp\alpha\right)}
\end{equation}
It should be noted that for small values of $\alpha$, the value of $L_0$ is smaller than the contact value of 2. For these situations, the value of $L_0$ loose its physical meaning, but acts as a way to extrapolate the approximation to the case $\alpha \to 0$.

\subsection{\label{sec_second virial coefficient}\emph{Second virial coefficient}}

The sticky electrostatic sphere pair potential is given by
 \begin{equation} \label{eqn:V}
    V\left(\rho\right) \approx
        \begin{cases}
        \infty & 0 < \rho < 2\\
        V_{M}(\rho)+V_{C} & 2 < \rho < 2+3\epsilon\\
        V_{SC}(\rho) & 2+3\epsilon < \rho < \infty \\
        \end{cases}
\end{equation}

Based on $\left(\ref{eqn:V}\right)$ the integral in $\left(\ref{eqn:B2*}\right)$ for computing $B_{2}^*$, can be simplified as a sum of various ranges.

\begin{equation}
\label{eqn:B2integrals}
    B_{2}^* = -\frac{3}{8}\left[ \int_{0}^{2}\left(e^{-V\left(\rho\right)}-1\right)\rho^2d\rho + \int_{2}^{2+3\epsilon}\left(e^{-V\left(\rho\right)}-1\right)\rho^2d\rho + \int_{2+3\epsilon}^{\infty}\left(e^{-V\left(\rho\right)}-1\right)\rho^2d\rho\right]
\end{equation}

Substituting equations $\left(\ref{eqn:morse}\right)$, $\left(\ref{eqn:tauEff}\right)$, $\left(\ref{eqn:Vsc|L_0}\right)$, $\left(\ref{eqn:L_0}\right)$ and $\left(\ref{eqn:V}\right)$ in $\left(\ref{eqn:B2integrals}\right)$:
\begin{equation}
\label{eqn:B2integralswithV}
    B_{2}^*= -\frac{3}{8}\left[ \int_{0}^{2}\left(e^{-\infty}-1\right)\rho^2d\rho + \int_{2}^{2+3\epsilon}\left(e^{-V_{M}(\rho)-V_{C}}-1\right)\rho^2d\rho + \int_{2+3\epsilon}^{\infty}\left(e^{-V_{SC}(\rho)}-1\right)\rho^2d\rho\right]
\end{equation}

\begin{equation}
\label{eqn:B2integralswithL_0}
    B_{2}^* \approx -\frac{3}{8}\left[ \frac{-8}{3} + e^{-V_C}\int_{2}^{2+3\epsilon}e^{-V_{M}}\rho^{2}d\rho + \int_{2+3\epsilon}^{L_0}\left(e^{-V_{SC}(\rho=L_0)}-1\right)\rho^2d\rho\right]
\end{equation}

\begin{equation} \label{eqn:B2*analytical}
    B_2^* \approx  1 - \frac{1}{4\tau^*}e^{-V_C}+b_{1}\left(\frac{e-1}{3e}\left(L_0\right)^3\right)
\end{equation}

Equation $\left(\ref{eqn:B2*analytical}\right)$ has a constant $b_{1}$ added to match the numerical data. The value of $b_{1}$  changes with the change in Debye length of the solution. The values of $b_1$ for $\kappa^{-1}=0.7,0.85,1,1.25,1.5$ are $b_1=0.6,0.78,0.93,1.05,1.1$ respectively.

\subsection{\label{sec_zero-shear viscosity}\emph{Zero-shear viscosity}}
Since the viscosity coefficient $c_2$ is also written as integrals in equation $\left(\ref{eqn:c2}\right)$, the integrals can be similarly approximated over the ranges $2<\rho<2+3\epsilon$ and $2+3\epsilon<\rho<\infty$. The integrals over the first range capture the contribution from particle separation distances where the sticky interactions play a direct role and the integrals over the second range yield the contribution in which the sticky interactions can only have an indirect role.

\subsubsection{\label{c2H}Hydrodynamic contribution}
The hydrodynamic contribution is given by
\begin{equation}
\label{eqn:c2H_SES}
  c_{2,SES}^{H} = \frac{5}{2}\left[1+3\left(\int_{2}^{2+3\epsilon}J\left(\rho\right)g\left(\rho\right)\rho^{2}d\rho+\int_{2+3\epsilon}^{\infty}J\left(\rho\right)g\left(\rho\right)\rho^{2}d\rho\right)\right]
 \end{equation}

Within the integrals, the function $g(\rho)$ is known analytically. For the function $J\left(\rho\right)$, Batchelor \textit{et~al.}~\cite{batchelor_1972b} provides a near-field value of 0.2214, intermediate tabulated values for the range $2+3\epsilon\leq\rho\leq3$, and a far field form $J\approx\frac{15}{2}\rho^{-6}$ as $\rho\to\infty$. Equation $\left(\ref{eqn:c2H_SES}\right)$ can hence be further split into:
\begin{equation}
c_{2,SES}^{H} = \frac{5}{2}\left[1+3\left(\int_{2}^{2+3\epsilon}J\left(\rho\right)g\left(\rho\right)\rho^{2}d\rho+\int_{2+3\epsilon}^{3}J\left(\rho\right)g\left(\rho\right)\rho^{2}d\rho+\int_{3}^{\infty}J\left(\rho\right)g\left(\rho\right)\rho^{2}d\rho\right)\right]
 \end{equation}
The first integral can be approximated as
\begin{equation}
\int_{2}^{2+3\epsilon}J\left(\rho\right)g\left(\rho\right)\rho^{2}d\rho
\approx J\left(2\right)e^{-V_C}\int_{2}^{2+3\epsilon}e^{-V_{M}}\rho^{2}d\rho \approx 0.2214e^{-V_C}\frac{2}{3\tau^{*}}
\end{equation}
For the range $2+3\epsilon<\rho<3$, we roughly approximate $V_{SC}\approx V_C$, in order to simplify $g(\rho)$, and then numerically integrate the tabulated data of $J$ to obtain
\begin{equation}
\int_{2+3\epsilon}^{3}J\left(\rho\right)g\left(\rho\right)\rho^{2}d\rho
\approx e^{-V_C}\int_{2+3\epsilon}^{3}J\left(\rho\right)\rho^{2}d\rho \approx 0.271e^{-V_C}
\end{equation}
For the last part, $3<\rho<\infty$, we can assume $g\left(\rho\right)\approx 1$ independent of $\alpha$ and the integral can be solved analytically as
\begin{equation}
\int_{3}^{\infty}J\left(\rho\right)g\left(\rho\right)\rho^{2}d\rho \approx
\frac{15}{2}\int_{3}^{\infty}\frac{1}{\rho^{4}}d\rho
\end{equation}
This approximation is sufficiently accurate for the values of $\alpha$ examined here. For larger $\alpha$, an alternative approximation, such as the excluded shell model, would be more accurate.

Combining the terms together gives
\begin{equation}
\label{eqn:c2H_SESanalytical}
 c_{2,SES}^{H}  = \frac{5}{2}+e^{-V_C}\left(\frac{1.11}{\tau^{*}}+ 2.03\right)+0.69
 \end{equation}

\subsubsection{\label{c2b}Brownian contribution}
The Brownian contribution is given by
\begin{equation}
\label{eqn:c2B_SES}
  c_{2,SES}^{B} = \frac{9}{40}\left[\left(\int_{2}^{2+3\epsilon}\rho^{2}W\left(\rho\right)g\left(\rho\right)f\left(\rho\right)d\rho+\int_{2+3\epsilon}^{\infty}\rho^{2}W\left(\rho\right)g\left(\rho\right)f\left(\rho\right)d\rho\right)\right]
 \end{equation}
This formula requires knowledge of the function $f(\rho)$, which is the result of solving a differential equation for each set of interaction parameters. In general, the $f$ resulting from the sum of two potentials is not the sum of the functions from each potential individually. In this work, the short ranged aspect of the Morse potential produces a $f$ that is nearly the sum of two contributions, one that depends on the Morse potential and one that depends on the screened Coulomb potential.

For the range $2<\rho<2+3\epsilon$, $W$ and $f$ are approximately constants. The $f$ at contact is denoted as $f_C$ and is approximated from the numerical results as
\begin{equation}
\label{eqn:f_C}
f_C \approx 1.37 + \frac{0.21\left(\frac{1}{\tau^*}\right) \ln{\epsilon }}{-0.05\left(\frac{1}{\tau^*}\right)\ln{\epsilon}+1} + \frac{2}{3}\alpha e^{-2\kappa}
\end{equation}
With these approximations, and the definition in equation (\ref{eqn:tauEff}), the first integral becomes
\begin{equation}
\int_{2}^{2+3\epsilon}\rho^{2}W g f d\rho \approx 6.372 e^{-V_C} f_C \int_{2}^{2+3\epsilon}\rho^{2}e^{-V_{M}}d\rho = 6.372 e^{-V_C} f_C \left(\frac{2}{3\tau^*}\right)
\end{equation}

For the contribution from the integral in the range $2+3\epsilon<\rho<\infty$, we can again ignore the dependence on $\alpha$ and use the hard-sphere result. Batchelor\cite{batchelor_1977} provides a value of 0.97 for the contribution from the second integral. Although this is not accurate at large $\alpha$, it is sufficient accurate at the relatively smaller $\alpha$ examined here.

Combining the two approximations gives the overall contribution of
\begin{equation}
\label{eqn:c2B_SESanalytical}
 c_{2,SES}^{B} = 0.96\left[\left(\frac{1}{\tau^*}\right)e^{-V_C}\left(1.37 + \frac{0.21\left(\frac{1}{\tau^*}\right) \ln{\epsilon }}{-0.05\left(\frac{1}{\tau^*}\right)\ln{\epsilon}+1} + \frac{2}{3}\alpha e^{-2\kappa}\right) + 1.01\right]
 \end{equation}

\subsubsection{\label{c2i}Interaction contribution}
The final contribution comes from the interactions directly, and is given by
\begin{equation}
\label{eqn:c2I_SES}
  c_{2,SES}^{I} = \frac{9}{40}\left[\left(\int_{2}^{2+3\epsilon}\rho^{3}\left(1-A\left(\rho\right)\right)\frac{dg}{d\rho}f\left(\rho\right)d\rho+\int_{2+3\epsilon}^{\infty}\rho^{3}\left(1-A\left(\rho\right)\right)\frac{dg}{d\rho}f\left(\rho\right)d\rho\right)\right]
 \end{equation}
Because of the derivative of $g$, the contribution from the first integral will be dominated by the separation $\rho \approx 2 + \epsilon$ where the Morse potential changes most rapidly. We can approximate $\frac{d g}{d \rho}$ as proportional to a Dirac delta function. The prefactor is chosen so that the approximation has the correct integral. That is, within the first integral we approximate
\begin{equation}
\frac{d g}{d \rho} \approx e^{-V_C} (1-e^{-\varepsilon_d}) \delta\left(\rho-\left(2+\epsilon\right)\right)
\end{equation}

Using this approximation and the function $A$ near contact, the contribution from the first integral is
\begin{align}
    &-\left(\frac{9}{40}\right)\left(8\right)e^{-V_C}\left(e^{\varepsilon_d}-1\right)\left(4.077\epsilon\right)\int_{2}^{2+3\epsilon}f\left(\rho\right)\delta\left(\rho-\left(2+\epsilon\right)\right)d\rho
    \end{align}
Finally, $f$ is approximated near contact and the definition of $\tau^*$ used to write the first contribution as
\begin{equation}
    -1.22\left(\frac{1}{\tau^*}\right)e^{-V_C}\left(1.37 + \frac{0.21\left(\frac{1}{\tau^*}\right) \ln{\epsilon }}{-0.05\left(\frac{1}{\tau^*}\right)\ln{\epsilon}+1} + \frac{2}{3}\alpha e^{-2\kappa}\right)
\end{equation}

For the range $2+3\epsilon<\rho<\infty$, the Morse potential does not strongly impact $g$ and we can ignore its impact on $f$, keeping only the impact from the electrostatic repulsions. For the second integral, we can follow previous literature with electrostatics alone and approximate these effects using excluded shell model for large $\alpha$.

Denoting the excluded volume radius as $L_{0}$ and substitution of the far-field forms of the hydrodynamic function into the second integral of $\left(\ref{eqn:c2I_SES}\right)$ produces a contribution of
\begin{equation}
\label{eqn:c2Iexcludedvol}
\frac{9}{40}L_{0}^{3}e^{\left( -\alpha\frac{e^{-\kappa L_0}}{ \kappa L_0}\right)}\frac{\alpha e^{-\kappa L_0}}{ L_0}\left[1-\frac{5}{L_{0}^{3}}+\frac{8}{L_{0}^{5}}\right]f\left(L_0\right)
\end{equation}
This expression requires an estimate of $f$ at the excluded shell radius.

We can estimate $f$ as
\begin{equation} \label{eqn:intermediaterangef}
    f\left(\rho\right) \approx\frac{1}{6}\rho^2\alpha e^{-\kappa\rho}+1.37
\end{equation}
This interpolates between the hard-sphere contact value of $1.37$ and an expression that balances electrostatic repulsions with Brownian motion, which is accurate at larger $\alpha$.

Substituting equation $\left(\ref{eqn:intermediaterangef}\right)$ at $\rho=L_{0}$ into equation $\left(\ref{eqn:c2Iexcludedvol}\right)$ gives a contribution from the second integral of
\begin{align}
  \frac{9}{40}L_{0}^{3}e^{\left(- \alpha\frac{e^{-\kappa L_0}}{ \kappa L_0}\right)}\frac{\alpha e^{-\kappa L_0}}{ L_0}\left[1-\frac{5}{L_{0}^{3}}+\frac{8}{L_{0}^{5}}\right]\left(\frac{1}{6}L_{0}^2\alpha e^{-\kappa L_{0}}+ 1.37\right)
  \end{align}
Finally, we can simplify for $L_0 > 1$ and using $\alpha\frac{e^{-\kappa L_0}}{\kappa  L_0}\approx 1$ to give
\begin{equation}
\label{eqn:c2ISES_SC}
   A_{1}\left(\frac{3}{40}\right)\left(\frac{1}{\kappa}\ln\frac{\alpha}{\ln\left(\alpha/\ln\alpha\right)}\right)^2{\ln\left(\frac{\alpha}{\ln\alpha}\right)}\left(\frac{1}{6}\left(\frac{1}{\kappa}\ln\frac{\alpha}{\ln\left(\alpha/\ln\alpha\right)}\right)^2{\ln\left(\frac{\alpha}{\ln\alpha}\right)}+1.37\right)
\end{equation}
The constant $A_{1}$ has been added to account for the approximations of the integral and match the analytical results quantitatively with the numerical solution. The value of $A_{1}$ changes with the change in Debye length of the solution. The values of $A_1$ for $\kappa^{-1}=0.7,0.85,1,1.25,1.5$ are $A_1=1.5,1.92,2.35,3,3.3$ respectively.

The overall interaction contribution combines the results of both integrals to give
\begin{equation}
\begin{split}
& c_{2,SES}^{I} = -1.22\left(\frac{1}{\tau^*}\right)e^{-V_C}\left(1.37 + \frac{0.21\left(\frac{1}{\tau^*}\right) \ln{\epsilon }}{-0.05\left(\frac{1}{\tau^*}\right)\ln{\epsilon}+1} + \frac{2}{3}\alpha e^{-2\kappa}\right)+ \\ & A_{1}\left(\frac{3}{40}\right)\left(\frac{1}{\kappa}\ln\frac{\alpha}{\ln\left(\alpha/\ln\alpha\right)}\right)^2{\ln\left(\frac{\alpha}{\ln\alpha}\right)}\left(\frac{1}{6}\left(\frac{1}{\kappa}\ln\frac{\alpha}{\ln\left(\alpha/\ln\alpha\right)}\right)^2{\ln\left(\frac{\alpha}{\ln\alpha}\right)}+1.37\right)
\end{split}
\end{equation}

\subsubsection{\label{c2total}Overall result}
\begin{equation}
\label{eqn:c2_SESanalytical}
    \begin{split}
        & c_{2,SES} = c_{2,SES}^{H} + c_{2,SES}^{B} + c_{2,SES}^{I} \\
        & c_{2,SES}^{H} = \frac{5}{2}+\left(e^{-V_C}\left(\frac{1.11}{\tau^{*}}+ 2.03\right)\right)+0.69 \\
         & c_{2,SES}^{B} = 0.96\left[\left(\frac{1}{\tau^*}\right)e^{-V_C}\left(1.37 + \frac{0.21\left(\frac{1}{\tau^*}\right) \ln{\epsilon }}{-0.05\left(\frac{1}{\tau^*}\right)\ln{\epsilon}+1} + \frac{2}{3}\alpha e^{-2\kappa}\right) + 1.01\right] \\
        & c_{2,SES}^{I} = -1.22\left(\frac{1}{\tau^*}\right)e^{-V_C}\left(1.37 + \frac{0.21\left(\frac{1}{\tau^*}\right) \ln{\epsilon }}{-0.05\left(\frac{1}{\tau^*}\right)\ln{\epsilon}+1} + \frac{2}{3}\alpha e^{-2\kappa}\right)+ \\ & A_{1}\left(\frac{3}{40}\right)\left(\frac{1}{\kappa}\ln\frac{\alpha}{\ln\left(\alpha/\ln\alpha\right)}\right)^2{\ln\left(\frac{\alpha}{\ln\alpha}\right)}\left(\frac{1}{6}\left(\frac{1}{\kappa}\ln\frac{\alpha}{\ln\left(\alpha/\ln\alpha\right)}\right)^2{\ln\left(\frac{\alpha}{\ln\alpha}\right)}+1.37\right)
    \end{split}
\end{equation}

\bibliographystyle{unsrt}

\begin{thebibliography}{10}

\bibitem{connolly_2012}
B.~D. Connolly, C.~Petry, B.~Yadav, S.~Demeule, N.~Ciaccio, J.~M.~R. Moore,
  S.~J. Shire, and Y.~R. Gokarn.
\newblock Weak interactions govern the viscosity of concentrated antibody
  solutions: High-throughput analysis using the diffusion interaction
  paramater.
\newblock {\em Biophy. J.}, 103:69--78, 2012.

\bibitem{zydney_2015}
E.~Binabaji, J.~Ma, and A.~L. Zydney.
\newblock Intermolecular interactions and the viscosity of highly concentrated
  monoclonal antibody solutions.
\newblock {\em Pharm. Res.}, 32:3102--3109, 2015.

\bibitem{ross_1977}
P.~D. Ross and A.~P. Minton.
\newblock Hard quasispherical model for the viscosity of hemoglobin solutions.
\newblock {\em Biochem. Biophys. Res. Commun.}, 76:971--976, 1977.

\bibitem{baxter_1968}
R.~J. Baxter.
\newblock Percus-yevick equation for hard spheres with surface adhesion.
\newblock {\em J. Chem. Phys.}, 49:2770--2774, 1968.

\bibitem{russel_1984}
W.~B. Russel.
\newblock The huggins coefficient as a mean for characterizing suspended
  particles.
\newblock {\em J. Chem. Soc. Faraday Trans. 2}, 80:31--41, 1984.

\bibitem{stell_1991}
G.~Stell.
\newblock Sticky spheres and related systems.
\newblock {\em J. Stat. Phys.}, 63:1203--1221, 1991.

\bibitem{holmes-cerfon_2013}
M.~Holmes-Cerfon, S.~J. Gortler, and M.~P. Brenner.
\newblock A geometrical approach to computing free-energy landscapes from
  short-ranged potentials.
\newblock {\em PNAS}, 110:E5--E14, 2012.

\bibitem{holmes-cerfon_condens}
M.~Holmes-Cerfon.
\newblock Sticky-sphere clusters.
\newblock {\em Annu. Rev. Condens. Matter Phys.}, 8:77--98, 2017.

\bibitem{holmes-cerfon_2017}
Y.~Kallus and M.~Holmes-Cerfon.
\newblock Free energy of singulr sticky-sphere clusters.
\newblock {\em Phys. Rev. E}, 95:1--18, 2017.

\bibitem{piazza_1998}
R.~Piazza, V.~Peyre, and V.~Degiorgio.
\newblock ``sticky hard spheres'' model of proteins near crystallization: A
  test based on the osmotic compressibility of lysozyme solutions.
\newblock {\em Phys. Rev. E}, 58:R2733--R2736, 1998.

\bibitem{lenhoff_2017}
L.~J. Quang, S.~I. Sandler, and A.~M. Lenhoff.
\newblock Anisotropic contributions to protein-protein interactions.
\newblock {\em J. Chem. Theory Comput.}, 10:835--845, 2014.

\bibitem{hayakawa_2016}
M.~Hayakawa, H.~Onoe, K.~H. Nagai, and M.~Takinoue.
\newblock Complex-shaped three-dimensional multi-compartmental microparticles
  generated by diffusional and marangoni microflows in centrifugally discharged
  droplets.
\newblock {\em Sci. Rep.}, 6:1--9, 2016.

\bibitem{fryling_1963}
C.~F. Fryling.
\newblock The viscosity of small particle, electrolye- and soap-deficient
  synthetic latex gels.
\newblock {\em J. Colloid Sci.}, 18:713--732, 1963.

\bibitem{prieve_1987}
D.~C Prieve and S.~G. Bike.
\newblock Electrokinetic repulsion between two charged bodies undergoing
  sliding motion.
\newblock {\em Chem. Eng. Comm.}, 55:149--164, 1987.

\bibitem{prieve_1990}
S.~G. Bike and D.~C. Prieve.
\newblock Electrohydrodynamic lubrication with thin double layers.
\newblock {\em J. Colloid Interface Sci}, 136:95--112, 1990.

\bibitem{carnero-ruiz_1997}
F.~J. Rubio-Hern\'{a}ndez, A.~I. G\'{o}mez-Merino, E.~Ruinz-Reina, and
  C.~Carnero-Ruiz.
\newblock The primary electroviscous effect of polystyrene latexes.
\newblock {\em Colloids Surf. A}, 140:295--298, 1997.

\bibitem{napper_1977}
D.~H. Napper.
\newblock Steric stabilization.
\newblock {\em J. Colloid Interface Sci}, 58:390--407, 1977.

\bibitem{morbidelli_2015}
L.~Nicoud, M.~Lattuada, A.~Yates, and M.~Morbidelli.
\newblock Impact of aggregate formation on the viscosity of protein solutions.
\newblock {\em Soft Matter}, 11:5513--5522, 2015.

\bibitem{feke_1984}
D.~W. Weaver and D.~L. Feke.
\newblock Ageing in colloidal flocs: Effect of double-layer relaxation.
\newblock {\em J. Colloid Interface Sci}, 103:267--283, 1985.

\bibitem{rogers_2005}
H.~R. Rogers, E.~Michel, C.~A. Bauer, S.~Vanderet, B.~K. Hansen, D.~Roberts,
  A.~Calvez, J.~B. Crews, K.~O. Lau, A.~Wood, D.~J. Pine, and P.~V. Schwartz.
\newblock Selective, controllable, and reversible aggregation of polystyrene
  latex microspheres via dna hybridization.
\newblock {\em Langmuir}, 21:5562--5569, 2005.

\bibitem{pathak_2013}
P.~S. Sarangapani, S.~D. Hudson, K.~B. Migler, and J.~A. Pathak.
\newblock The limitations of an exclusively colloidal view of protein solution
  hydrodynamics and rheology.
\newblock {\em Biophys. J.}, 105:2418--2426, 2013.

\bibitem{pathak_2015}
P.~S. Sarangapani, S.~D. Hudson, R.~L. Jones, J.~F. Douglas, and J.~A. Pathack.
\newblock Critical examination of the colloidal particle model of globular
  proteins.
\newblock {\em Biophys. J.}, 108:724--737, 2015.

\bibitem{swan_2019}
G.~Wang, A.~M. Fiore, and J.~W. Swan.
\newblock On the viscosity of adhesive hard sphere dispersions: Critical
  scaling and the role of rigid contacts.
\newblock {\em J. Rheo.}, 63:229--245, 2019.

\bibitem{batchelor_1977}
G.~K. Batchelor.
\newblock The effect of brownian motion on the bulk stress in a suspension of
  spherical particles.
\newblock {\em J. Fluid Mech.}, 83(1):97--117, 1977.

\bibitem{mewis_2011}
J.~Mewis and N.~J. Wagner.
\newblock {\em Colloidal Suspension Rheology}.
\newblock Cambridge University Press, 2011.

\bibitem{russel_1976}
W.~B. Russel.
\newblock The rheology of suspensions of charged rigid spheres.
\newblock {\em J. Fluid Mech.}, 85:209--232, 1978.

\bibitem{neal_1998}
B.~L. Neal, D.~Asthagiri, and A.~M. Lenhoff.
\newblock Molecular origins of osmotic second virial coeffcieints of proteins.
\newblock {\em Biophys. J.}, 75:2469--2477, 1998.

\bibitem{batchelor_1972b}
G.~K. Batchelor and J.~T. Green.
\newblock The determination of the bulk stress in a suspension of spherical
  particles to order $c^2$.
\newblock {\em J. Fluid Mech.}, 56(3):401--427, 1972.

\bibitem{batchelor_1976}
G.~K. Batchelor.
\newblock Brownian diffusion of particles with hydrodynamic interaction.
\newblock {\em J. Fluid Mech.}, 74(1):1--29, 1976.

\bibitem{russel_1980}
W.~B. Russel.
\newblock Review of the role of colloidal forces in the rheology of
  suspensions.
\newblock {\em J. Rheo.}, 24:287--317, 1980.

\bibitem{russel_book_1987}
W.~B. Russel.
\newblock {\em The Dynamics of Colloidal Systems}.
\newblock The University of Wisconsin Press, Madison, 1987.

\bibitem{cichocki_1993}
B.~Cichocki, B.~U. Felderhof, K.~Hinsen, E.~Wajnryb, and J.~Blawzdziewicz.
\newblock Friction and mobility of many spheres in stokes flow.
\newblock {\em J. Chem. Phys.}, 100(5):3780--3790, 1994.

\bibitem{bergenholtz_2002}
J.~Bergenholtz, J.~F. Brady, and M.~Vicic.
\newblock The non-newtonian rheology of dilute colloidal suspensions.
\newblock {\em J. Fluid Mech.}, 456:239--275, 2002.

\bibitem{batchelor_1972}
G.~K. Batchelor and J.~T. Green.
\newblock The hydrodynamic interaction of two small freely-moving spheres in a
  linear flow field.
\newblock {\em J. Fluid Mech.}, 56(2):375--400, 1972.

\bibitem{cichocki_1990}
B.~Cichocki and B.~U. Felderhof.
\newblock Diffusion coefficients and effective viscosity of suspensions of
  sticky hard spheres with hydrodynamic interactions.
\newblock {\em J. Chem. Phys.}, 93:4427--4432, 1990.

\bibitem{cichocki_1988_diff}
B.~Cichocki and B.~U. Felderhof.
\newblock Short-time diffusion coefficients and high frequency viscosity of
  dilute suspensions of spherical brownian particles.
\newblock {\em J. Chem. Phys.}, 89:1049--1054, 1988.

\bibitem{cichocki_1988_susp}
B.~Cichocki and B.~U. Felderhof.
\newblock Long-time self-diffusion coefficient and zero-frequency viscosity of
  dilute suspensions of spherical brownian particles.
\newblock {\em J. Chem. Phys.}, 89:3705--3709, 1988.

\bibitem{bergenholtz_1994}
J.~Bergenholtz and N.~J. Wagner.
\newblock The huggins coefficient for the square-well colloidal fluid.
\newblock {\em Ind. Eng. Chem. Res.}, 33(10):2391--2397, 1994.

\bibitem{khair_2006}
A.~S. Khair and J.~F. Brady.
\newblock Single particle motion in colloidal dispersions: a simple model for
  active and nonlinear microrheology.
\newblock {\em J. Fluid Mech.}, 557:73--117, 2006.

\end{thebibliography}

\providecommand{\noopsort}[1]{}\providecommand{\singleletter}[1]{#1}%

\end{document}